# A Revision of the Bernoulli Equation as a Controller of the Fick's Diffusion Equation in Drug Delivery Modeling

**Ali Esmaeili [1], Saeed Ranjbar [2*]**

1- Department of Biotechnology, Science and Research Branch, Islamic Azad University, Tehran, Iran
E-mail: ali.esmaeili@srbiau.ac.ir

2 *- Department of Cardio-Thoracic of surgery, Maastricht University of Medical Center, Maastricht, Netherlands
E-mail: saeed.ranjbar@mumc.nl

**Abstract:**

Mathematical equations can be used as effectual tools in drug delivery systems modeling and are also highly helpful to have a theoretical understanding of controlled drug release and diffusion mechanisms. In this study we aim to present a mathematical combination between the Bernoulli equation and the Fick's equation as a diffusion controller in drug delivery systems. For this propose we have revised the Bernoulli equation as an additional, controller and complementary method of the Fick's diffusion equation to detect the optimal delivery direction to control the diffusion divergence of the drug carrier in vascular systems during the transportation process in biological tissues. Therefore, by utilizing the Bernoulli equation we could determine the real direction by the route function "f".

**Keywords:** Drug Delivery System, Fick's Laws of Diffusion, Bernoulli Equation, Mathematical Modeling.

## 1. Introduction:

By roughly speaking, to investigate a phenomenon in nature, a mathematical modeling with the design of mathematical equations based on the physical laws governing that phenomenon, gives an estimation of the behavior of that natural phenomenon. Here the natural phenomenon is a drug delivery system. Drug delivery process is generally defined as the utilization of methods and technologies for carrying the pharmaceutical substances in different routes of the body with a specific concentration to the certain tissue [1]. Therefore, due to the importance of drug delivery system in the treatment of human ailments, the mathematical models and equations have significant roles to gain a good understanding of the theoretical mechanisms behind empirical studies in drug delivery modeling development. Hence, to improve the effectiveness of the drug delivery systems, it's needed to consider the characterizations of the drug or drug carriers, physiological conditions of the body and physiochemical impacts of bloodstream. In this purpose, mathematical equations are helpful to formulate and design the desired drug delivery systems to predict and detect the release and diffusion mechanisms of drug carrier. Nevertheless, it's imperative to specify the drug carrier movement direction and its concentration changes in vascular system and blood flow to estimate the drug diffusion rate during the motion to minimize the release or diffusion out under the physiological and transportation conditions that the desired volume of the drug leads to the target tissue to improve the therapeutic efficacy.

The diffusion phenomenon can be generally described as the diffuse or random motion of molecules with a concentration gradient. This process is also important in the metabolism and exchange of substances between cells and bloodstream [2]. In the Fick's equation, different features such as the size and geometry of carrier, device structure and mass concentration are considered as the parameters to determine the diffusion coefficient in drug delivery systems modeling. Commonly **Fick's laws of diffusion** are used in drug delivery systems studies which can be described by following formulas:

**Fick's first law:**

$$F = -D\frac{\partial c}{\partial x}$$

Here $F$ is the rate of transfer or flux; $c$ is the concentration of the diffusing species and $D$ is diffusion coefficient or diffusivity.

**Fick's second law:**

$$\frac{\partial c}{\partial x} = D\left(\frac{\partial^2 c}{\partial x^2} + \frac{\partial^2 c}{\partial y^2} + \frac{\partial^2 c}{\partial z^2}\right)$$

Here $c$ is the concentration of diffusing species; $t$ is the time; $D$ is diffusion coefficient and $x, y$ and $z$ are the spatial coordinates. [3, 4]

The objective of this paper is to revise the **Bernoulli equation** (In fluid dynamics, The Bernoulli Equation states the changes of fluid pressure in regions where the flow velocity is altered) and combine it with Fick's diffusion equation as a controller and third law of diffusion in drug delivery systems modeling. This combination might be a helpful suggestion in drug diffusion controlling system to decrease the drug diffusion in biological routes by minimizing the drug carrier divergence in optimal detected direction where causes to prevent the diffusion of the drug out from the carrier during the transportation process in vascular systems.

2. **Method and Result:**

The problem is to determine the shape "f(y)" of a wire equation notch (here the notch is considered a partial origin of the drug bound), in which the volume flow rate of fluid, Q, through the partial origin of the drug bound is expressed as a function of height h of the mentioned origin. We first establish the equation:

$$\boldsymbol{Q(h) = c\int_0^h (h-y)^{\frac{1}{2}} f(y) dy}.$$

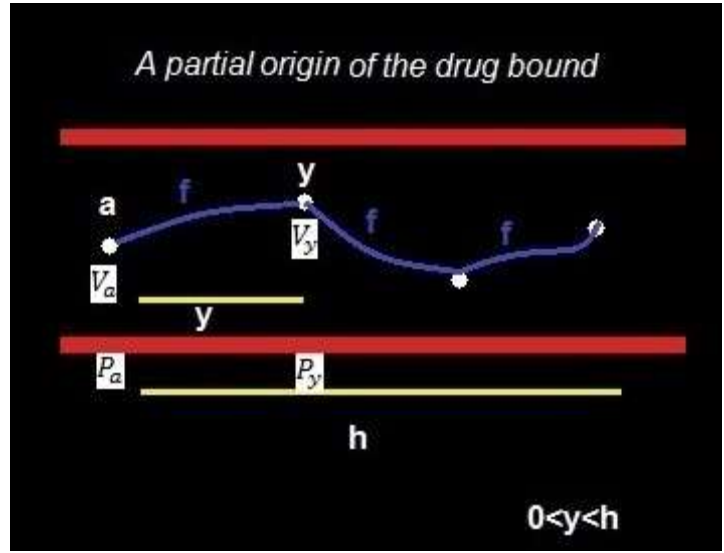

Figure 1: Consider front views of the partial origin of the drug bound:

Assuming Bernoulli's equation can be applied between the point "a" and "y" in Figure 1 [5,6], we obtain

(*) $$\frac{P_a}{\rho} + gh + \frac{V_a^2}{2} = \frac{P_y}{\rho} + gy + \frac{V_y^2}{2}.$$

Where $P_a$, $P_y$, $V_a$, $V_y$ are the pressures and velocities at point "a" and "y"; g is the graviational acceleration and ρ is the fluid density.

The pressures at "a" and "y" are both taken to be nearly atmospheric so that $\mathbf{P_a = P_y}$ and the velocity at "a" is assumed to be negligible ($V_a = 0$) since the fluid behind the origin is slow moving. Thus, (*) becomes

$$gh = gy + \frac{V_y^2}{2}$$

**So that**

$$V_y = (2g)^{1/2}(h-y)^{1/2}$$

Gives the velocity of the fluid at distance "y" through the partial origin of the drug bound in Figure 1.

The elemental area (Figure 1) is given by:

$$dA = 2f(y)dy.$$

So, by definition, the elemental volume flow rate through dA is

(**) $$dQ = V_y dA = 2(2g)^{\frac{1}{2}}(h-y)^{\frac{1}{2}}f(y)dy.$$

Denoting $2(2g)^{\frac{1}{2}}$ by c and integrating (**) from y = 0 to y = h gives the total volume flow rate through the partial origin of the drug bound:

$$Q(h) = c \int_0^h (h-y)^{\frac{1}{2}} f(y) dy$$

We find f(y) by finding f(h) inversely by the following formula:

$$f(h) = 1/\Gamma\left(\frac{3}{2}\right) \times \frac{d^2}{dh^2} \times 1/\Gamma\left(\frac{1}{2}\right) \times \int_0^h (h-y)^{-\frac{1}{2}} g(y) dy$$

Where $\frac{Q(h)}{c} = g(h)$ and $\Gamma$ is the Gama function. Since $\frac{Q(h)}{c}$ is known g(h) and g(y) are known. For the partial origin of the drug bound with numerical calculations, we can consider $g(h) = h^a$ approximately. For example, for a=7/2 is shaped like parabola for the partial origin of the drug bound. Now let "C" is the solution of the Fick equations, the main part of this paper is to consider the convolution between f and C, $f \otimes C$. This convolution means that the drug moves in the real direction by the rout function **f** (Figure 1) and the diffusion was released at the real direction into the tissue.

## 3. Discussion:

According to the mathematical equation, then solving this equation and with revise in the Bernoulli equation which is so valuable in drug delivery system context, we could predict the release and diffusion direction. With respect to the importance of the problem, too many studies have been conducted in this research area. for example, Khanday et al. 2017 [7], used three mathematical models based on the Fick's law of diffusion and law of mass action to study the drug diffusion in biological systems. They considered the blood and tissue as the compartments and studied the concentration changes in these compartments and revealed that the drug concentration level is reduced in one compartment and increased in other compartment over the time toward the tissue. Also, they could utilize the law of mass action and the Fick's law of diffusion in normal situations to understand the drug concentration changes in compartments over the time. But the advantageous of there is when the law of mass action is applicable that the drug mass would not be eliminated by blood cells during the transportation process. As another example, Chakravarty and Dalal. 2016 [8], assumed that the mass transform is in the normal direction to the tissue and then they followed three steps in the work, 1, drug dynamics in the polymeric matrix, 2, drug direction in the biological tissue along with initial, interface and boundary conditions assumptions and 3, mathematical solution of partial differential equations which are governed in the step 1 and step 2. But in comparison with our study, we don't consider the linear normal direction to the tissue, but using Bernoulli equation we have the real direction by the route function **f** (Figure 1). In our method the combination of Bernoulli equation with Fick's equation is completely independent from the type of the disease and patient and it's completely in accordance with the physiological conditions, vascular systems and blood flows.

## 4. Conclusion:

In this paper we successfully developed an optimized mathematical model for controlled drug delivery in bio-tissues. So, it has to be pointed out that the revised Bernoulli equation was needed to be considered and joined to the Fick's laws of diffusion as a complementary method for specifying the optimal and real delivery direction and leads the drug carrier into this direction where the drug diffusion is controlled with lowest divergence during the motion in the vascular systems into the tissues.

**There is no conflict of interest.**